\documentclass[12pt,preprint]{aastex}
\shorttitle{Correlations of disk and jet emission}
\shortauthors{Lin et al.}
\begin{document}
\title{Correlations of Disk and Jet Emission Deviating from the Fundamental Plane}

\author{Da-Bin Lin\altaffilmark{1, 2}, Wei-Min Gu\altaffilmark{3}, Hui-Jun Mu\altaffilmark{1, 2}, Zu-Jia Lu\altaffilmark{1, 2}, Ren-Yi Ma\altaffilmark{3}, En-Wei Liang\altaffilmark{1, 2}}
\altaffiltext{1}{GXU-NAOC Center for Astrophysics and Space Sciences， Department of Physics, Guangxi University, Nanning 530004, China; lindabin@gxu.edu.cn}
\altaffiltext{2}{Guangxi Key Laboratory for Relativistic Astrophysics, the Department of Physics, Guangxi University, Nanning 530004, China;}
\altaffiltext{3}{Department of Astronomy and Institute of Theoretical Physics and Astrophysics, Xiamen University, Xiamen, Fujian 361005, China}
%%%%%%%%%%%%%%%%%%%%%%%%%%%%%%%%%%%%%%%%%%%%%%%%%%%%%%%%%%%%%%%%%%%%%%%%%%%%%%%%%%%%%%%%%%%%%%%%%%%%%%%%%%%%%%%%%
%%%%%%%%%%%%%%%%%%%%%%%%%%%%%%%%%%%%%%%%%%%%%%%%%%%%%%%%%%%%%%%%%%%%%%%%%%%%%%%%%%%%%%%%%%%%%%%%%%%%%%%%%%%%%%%%%
\begin{abstract}
The variability of accretion rate,
which is believed to induce the aperiodic variability of X-ray emission from disk,
may affect the energy injection into the jet.
In this spirit,
a correlation between disk emission and jet emission can be formed
even if the mean luminosity of disk emission keeps constant.
In this work, these correlations are found
in the situation that the luminosity of disk emission is variable and kept with a constant mean value.
The obtained correlations may be shallower than that of the fundamental plane of black hole activity.
In addition, the slope of correlation may increase with increasing observed frequency of jet emission.
For the luminosities spacing with three days, the slope of correlation decreases with increasing black hole mass.
The deviation of our found correlations from that of the fundamental plane
is related to the suppression of variability in the jet emission
in comparison with that in the disk emission.
This mechanism may work in some sources in which shallower correlations have been reported.
Moreover, it implies that luminosities used to estimate the relation of fundamental plane should cover an appropriate timescale,
in which the variability of jet emission is not significantly suppressed.
\end{abstract}
%%%%%%%%%%%%%%%%%%%%%%%%%%%%%%%%%%%%%%%%%%%%%%%%%%%%%%%%%%%%%%%%%%%%%%%%%%%%%%%%%%%%%%%%%%%%%%%%%%%%%%%%%%%%%%%%%
\section{Introduction} \label{sec:Introduction}
The jet produced in an accreting black hole seems to be universal,
since it is widely found in active galactic nuclei (AGNs) and Galactic black hole binaries (BHBs).
The emission from jets and accretion disks can be observed in the energy bands from radio to X-ray.
For accretion systems with a jet,
it is believed that the radio emission is
mainly produced in the jet by the synchrotron process (\citealp{Begelman1984}).
However, the X-ray emission is usually associated with the inner region of disk,
where the temperature is highest and the corona may be formed (\citealp{Liang1977}; \citealp{Haardt1993}).
Since the jet is directly related to the accreting process,
a correlation between radio and X-ray luminosities can be naturally expected.
The existence of a tight, non-linear correlation between the radio ($L_{\rm R}$) and
X-ray luminaries $(L_{\rm X})$, i.e., $L_{\rm R}\propto L_{\rm X}^{0.7\pm0.1}$,
was found in BHBs during the low/hard state (\citealp{Corbel2003}; \citealp{Gallo2003}).
This correlation was quickly extended to include AGNs,
which is powered by accretion onto a super-massive black hole.
The relation is the so-called ``fundamental plane'' of black hole activity,
which can be described as ${\rm log}L_{\rm R} = (0.60_{-0.11}^{+0.11}){\rm log}L_{\rm X} + (0.78_{-0.09}^{+0.11}){\rm log}(M_{\rm BH}/M_\odot)+7.33$ (\citealp{Merloni2003}; \citealp{Falcke2004}; \citealp{Merloni2006}).
Here, $L_{\rm R}$ is the nuclear radio luminosity at frequency $\nu = 5\;\rm GHz$,
$L_{\rm X}$ is the nuclear X-ray luminosity at $2-10\;\rm keV$,
$M_{\rm BH}$ is the black hole mass,
and $M_\odot$ is the solar mass.
Such a plane confirms that the production of jets (traced by radio luminosity) is fundamentally
associated with the accreting process (traced by X-ray luminosity).

The emission from accretion disks in BHBs and AGNs displays significant aperiodic variabilities
on a broad range of timescales.
The most promising explanation for these variabilities
is the model of propagating fluctuations
(\citealp{Lyubarskii1997}; \citealp{King2004}; \citealp{Mayer2006}; \citealp{Janiuk2007}; \citealp{Lin2012}).
In this model, the variabilities of disk emission is induced by the fluctuations of accretion rate,
which are caused at different radii and propagate into the inner region of the disk.
The fluctuations of accretion rate in the inner region would
modulate the energy released in the vicinity of black hole, where produces most of X-ray emission.
Since the production of jets is fundamentally associated with the accreting process,
the energy injected into the jet should be affected by the fluctuations of accretion rate.
In this work, we present the variability of jet power and thus the jet emission
in the model of propagating fluctuations.
The effects of these variabilities on the relation of fundamental plane are our main focus.

This paper is organized as follows.
The models for variability of disk emission and jet emission are presented in Section~\ref{sec:Model}.
The light curves produced based on the above models are shown in Section~\ref{sec:Result}.
Our main focus in the present work, i.e.,
the relations between disk emission and jet emission, are also discussed in this section.
In Section~\ref{sec:Conclusion}, we summary conclusions and present a brief discussion.

\section{Models}\label{sec:Model}
\subsection{Variability of accretion disk emission}\label{sec:Disk Variability}
The model of propagating fluctuations in mass accretion rate is proposed to explain
the complex variability of emission from accreting disk
(\citealp{Lyubarskii1997}; \citealp{Kotov2001}; \citealp{Arevalo2006}).
In this picture, fluctuations stirred up far from the black hole modulate
the mass accretion rate and thus the emission from the disk, which is mainly produced in the inner region of the disk.
The fluctuations are produced on timescales related to the viscous timescale at the
radius of its origin and are uncorrelated for different radial scales.
In this scenario, the mixture of fluctuations with different timescales is a multiplicative process.
Then, the emitted flux from accretion disk is naturally predicted to display a linear rms~-~flux relation,
which is observed in a number of AGNs and BHBs (\citealp{Uttley2001}; \citealp{Uttley2005}; \citealp{Heil2012}).

To generate light curves according to the model of propagating fluctuations,
the energy release profile of accretion flow in the radial direction,
which determines the emission behavior of disk, should be assumed.
In this paper, $\epsilon(r) = r^{-3}(1-\sqrt{r_{\rm in}/r})$ is taken
to follow the radial loss rate of gravitational energy in the accretion disk,
where $r$ represents the radius, $r_{\rm in}$ is the inner radius of the disk
and is fixed at $3 r_{\rm g}$ in the following simulations,
and $r_{\rm g}=2GM_{\rm BH}/{c^2}$ is the Schwarzschild radius of black hole.
In the standard thin accretion disc (\citealp{Shakura1973}), the emitted flux per unit area, $F(r,t)$,
is proportional to $\epsilon(r)$ and $\dot{M}(r,t)$, i.e., $F(r,t)\propto \dot{M}(r,t)\epsilon(r)$,
where $\dot{M}(r,t)$ is the mass accretion rate at the radius $r$ and time~$t$.
In this situation, the luminosity of disk emission can be read as
\begin{equation}\label{EQ:L_X_initial}
L_{\rm X}
=
\int_{r_{\rm in}}^{r_{\rm{out}}}{F(r,t)2\pi r dr}
\propto
\int_{r_{\rm in}}^{r_{\rm{out}}}{\dot{M}(r, t)\epsilon(r)2\pi r dr}.
\end{equation}
As pointed out by the \cite{Arevalo2006}, the emissivity profile of different photons may be different
since the emitted spectrum might be radius-dependent.
Then, we should modify the function of $\epsilon(r)$ to describe the emissivity profile of different photons.
For this purpose, an adequate value of $\gamma$ can be introduced and
the emissivity profile is modified as $\epsilon(r, \gamma) = r^{-\gamma}(1-\sqrt{r_{\rm in}/r})$ for different photons.
To our knowledge, the X-ray emission may be produced in an accreting corona (\citealp{Liang1977}),
possibly sandwiching an optically thick, geometrically
thin disk, or in the advection-dominated accretion flow (ADAF: \citealp{Narayan1994}; \citealp{Yuan2014}).
In the ADAF,
which is the focused accretion flow in this work (see the discussion in Section~\ref{Sec:Variability of Jet Power}),
the emitted flux per unit area $F(r,t)$ is proportional to $M(r, t)^2$
rather than $\dot{M}(r, t)$.
Then, Equation~(\ref{EQ:L_X_initial}) can be turned into
\begin{equation}\label{EQ:L_X}
L_{\rm X}
\propto
\int_{r_{\rm in}}^{r_{\rm{out}}}{\dot{M}(r, t)^2\epsilon(r,\gamma)r dr}
\end{equation}
in the ADAF.
This is the equation used to produce the light curves of X-ray emission from disk in this work.
It is shown that the light curves produced with different $\gamma$ ($\gamma > 2.5$) does not show significant difference
except for the variability in the shortest timescales (see Figure~2 of \citealp{Arevalo2006}),
which does not affect the relations studied in this work (please see the discussion in Section~\ref{sec:Analysis}).
Then, we discuss the fluctuations of disk emission with $\gamma=3$.

\subsection{Variability of Jet Power}\label{Sec:Variability of Jet Power}
The dominant paradigms for jet production are outlined in the works of \cite{Blandford1977}
(BZ model) and \cite{Blandford1982} (BP model).
In both of the models, the large-scale ordered magnetic fields threading the accretion flow
or spinning black hole are required.
It is suggested that the large-scale magnetic fields may diffuse outward rapidly
if the half-thickness of the accretion disk $H$ is significantly less than the radius $r$
(\citealp{Lubow1994}; \citealp{Heyvaerts1996}; \citealp{Guan2009}).
Then, the jet discussed in our paper should be produced in a geometrically thick disk
(\citealp{Maccarone2003}; \citealp{Sikora2007}; \citealp{Coriat2011}; \citealp{Russell2011}),
such as the ADAF or the slim disk (\citealp{Abramowicz1988}).
Our work focuses on the jet produced in the ADAF, of which the accretion rate is relatively low.
The jet power ($P_{\rm jet}$) in the BP and BZ models is associated
with the strength of large-scale ordered poloidal magnetic field $B_{\rm p}$.
In the BP model, the jet power can be described as (e.g., \citealp{Livio1999}; \citealp{Cao2002}; \citealp{Li2012})
\begin{equation}\label{EQ:Jet Energy_01}
{P_{\rm jet}
\sim \int_{r_{\rm in} }^{r_{\rm{out}} } {\Omega \left(r{\frac{B_{\rm p}^2 }{4\pi }} \right)w(r)2\pi rdr}},
\end{equation}
where $\Omega(r)\propto r^{-3/2}$ is the angular velocity of disk,
$rB_{\rm p}^2/(4\pi)$ is the torque applied on the disk, and
$w(r) \propto r^{-\mu}$ is introduced to describe the contribution of different radius to the jet power.
In the BP model, the energy of jet is mainly from the inner region of disk.
Then, $\mu$ should be larger than or equal to $0$, i.e.,
\begin{equation}
\mu\gtrsim 0.
\end{equation}
The strength of large-scale poloidal magnetic field
remains uncertain, but is believed to be associated with the accreting process
(e.g., \citealp{Moderski1996}; \citealp{Ghosh1997}; \citealp{Livio1999}; \citealp{Meier2001}; \citealp{Nemmen2007}).
In addition, the strength of small-scale magnetic field ($B_{\rm{p},d}$) in the magnetohydrodynamic turbulence of the disk
is usually used to estimate the strength of $B_{\rm p}$, i.e.,
$B_{\rm p} \sim B_{\rm{p},d}$ in the ADAF (\citealp{Livio1999}; \citealp{Meier2001}; \citealp{Wu2008}).
In the accretion flow, $B_{\rm p}$ may be radius-dependent.
In order to model this behavior, we adopt (\citealp{Livio1999}; \citealp{Meier2001}; \citealp{Wu2008}; \citealp{Li2012})
\begin{equation}\label{EQ:B_p}
B_{\rm p}^2 \propto \tau _{r\phi } (r/3r_{\rm g})^{-s},
\end{equation}
where $\tau _{r\phi }={\left\langle {\delta {B_r}\delta {B_\varphi }/4\pi - \rho \delta {\upsilon _r}\delta {\upsilon _\varphi }} \right\rangle _\varphi } \propto B_{\rm{p},d}^2$
is the stress in the magnetohydrodynamic accretion flow.
If the parameter $s$ in Equation (\ref{EQ:B_p}) is larger than $0$, $B_{\rm p}$ in the outer region will suppress
the magnetorotational instability (\citealp{Velikhov1959}; \citealp{Chandrasekhar1960}; \citealp{Balbus1991}, 1998),
which is responsible for the angular momentum transport.
Then, $s$ should be larger than or equal to $0$, i.e.,
\begin{equation}
s \gtrsim 0.
\end{equation}
Based on the above discussion, Equation (\ref{EQ:Jet Energy_01}) becomes
\begin{equation}\label{EQ:Jet Energy}
P_{\rm jet}(t)
=
\zeta\int_{{r_{{\rm{in}}}}}^{{r_{{\rm{out}}}}} {\dot{M}(r,t)r^{-\mu-s}{\epsilon}(r, 3)rdr}
=
\zeta\int_{{r_{{\rm{in}}}}}^{{r_{{\rm{out}}}}} {\dot{M}(r,t){\epsilon}(r, 3+\mu+s)rdr},
\end{equation}
where $H\sim r$ and $\Omega\tau_{r\varphi}H\propto\dot{M}{\epsilon}(r,3)$ are taken and $\zeta$ is a constant.
In the BZ model, the variability of jet power,
which is induced by the fluctuations of magnetic field in the vicinity of black hole,
does not present significant difference with Equation (\ref{EQ:Jet Energy}).
Then, we use Equation (\ref{EQ:Jet Energy}) to describe the jet power.
This equation reveals that the power of jet depends linearly on the accretion rate in the ADAF,
which has been used in a number of works (e.g., \citealp{Meier2001}; \citealp{Nemmen2007}; \citealp{MartinezSansigre2011}).
Equation (\ref{EQ:Jet Energy}) is similar to Equation~(\ref{EQ:L_X_initial})
but with $\gamma=3+\mu+s$ and $\mu+s \gtrsim 0$.
It was discussed previously that
the variabilities based on Equation~(\ref{EQ:L_X}) (or Equation~(\ref{EQ:Jet Energy}))
with varying $\gamma$ ($\gamma > 2.5$) do not show significant difference.
Then, the variability of jet power based on Equation (\ref{EQ:Jet Energy})
would follow that of accretion rate or $L_X$ except for the amplitude of variability,
which has been found in the observation of GX~$339-4$ (\citealp{Casella2010}).
This fact implies that Equation (\ref{EQ:Jet Energy}) is applicable to describe the jet power.
In our work, we take $\mu+s=0$ (please see the discussion in Section~\ref{sec:Analysis}).
In this situation, we take $\zeta=0.1$,
which corresponds to the case that $10\%$ of gravitational energy released in the accretion process enters the jet.

In the following part of this Section, we try to work out the variability of jet emission
based on the jet power presented in Equation (\ref{EQ:Jet Energy}).
The jet emission is modeled in the internal shock scenario,
which has been widely adopted to explain the broadband
spectral energy distribution of AGNs, BHBs, and the prompt emission of gamma-ray bursts
(\citealp{Piran1999}; \citealp{Spada2001}; \citealp{Yuan2005}; \citealp{Wu2007}; \citealp{Nemmen2011}; \citealp{Sun2013}).
Accompanying with the accretion onto a black hole,
a small fraction of the accreted material $\dot{M}_{\rm jet}(t,z=0)=P_{\rm jet}(t)/(\overline{\Gamma}-1)c^2$
is transferred into the vertical direction,
where $z$ represents the distance measured along the jet with respect to the black hole.
With the propagation of $\dot{M}_{\rm jet}$ in this direction, the phenomena of jet are formed.
Since $P_{\rm jet}(t)$ varies with time,
the value of $\dot{M}_{\rm jet}(t, z)(=P_{\rm jet}(t-z/\upsilon_{\rm jet})/(\overline{\Gamma}-1)c^2)$
may be different for different $z$ at the same time $t$,
where ${\upsilon _{{\rm{jet}}}} = c(1 - {\overline{\Gamma} ^{ - 2}})^{1/2}$ is the velocity of jet
and $c$ is the velocity of light.
According to the internal shock model,
the Lorentz factor of jet varies with time, and as a result,
faster portions of the jet catch up with slower portions of the jet.
The internal shocks are formed in the ensuing collision, which occurs in the region of $z\geqslant z_0$.
In the shocks, a small fraction $\xi_e$ of the electrons
are accelerated and form a power-law energy distribution with index $p$.
The broad-band emission of jet is mainly produced by these electrons through synchrotron process
(\citealp{Markoff2001}).
In order to calculate the jet emission, the jet half-opening angle $\phi$,
the average bulk Lorentz factor $\overline{\Gamma}$ of the jet,
the angle $\psi$ of the jet with respect to the line of sight,
and the energy density ratio  $\epsilon_e$ ($\epsilon_B$) of accelerated electrons
(amplified magnetic field) to the shock energy
should be prescribed.
In the present work, the calculation of jet emission follows the procedure
presented in the Appendix of \cite{Yuan2005}.
The values of $\phi= 0.1\;\rm radians$, $\overline{\Gamma} = 1.2$, $\psi=30^{^\circ}$,
$\xi_e=1\%$, $p=2.1$, $\epsilon_e=0.06$, and $\epsilon_B=0.02$ are adopted in our calculation.
The average of $\dot{M}$ is taken as $1.39\times 10^{16} (M_{\rm BH}/M_{\odot})\;{\rm g/s}=0.01\dot{M}_{\rm Edd}$,
i.e., $\overline{\dot{M}}=0.01\dot{M}_{\rm Edd}$, where $\dot{M}_{\rm Edd}$ is the Eddington accretion rate.
In this paper, we focus on the variability of jet emission induced
by the fluctuation of accretion rate, i.e., Equation (\ref{EQ:Jet Energy}).
Then, the variability due to the collisions between shells in the internal shock model,
which may increase the scattering of our studied correlations,
is not considered in this work.

\section{Relations between disk emission and jet emission}\label{sec:Result}
\subsection{Results}
\begin{figure}
\plotone{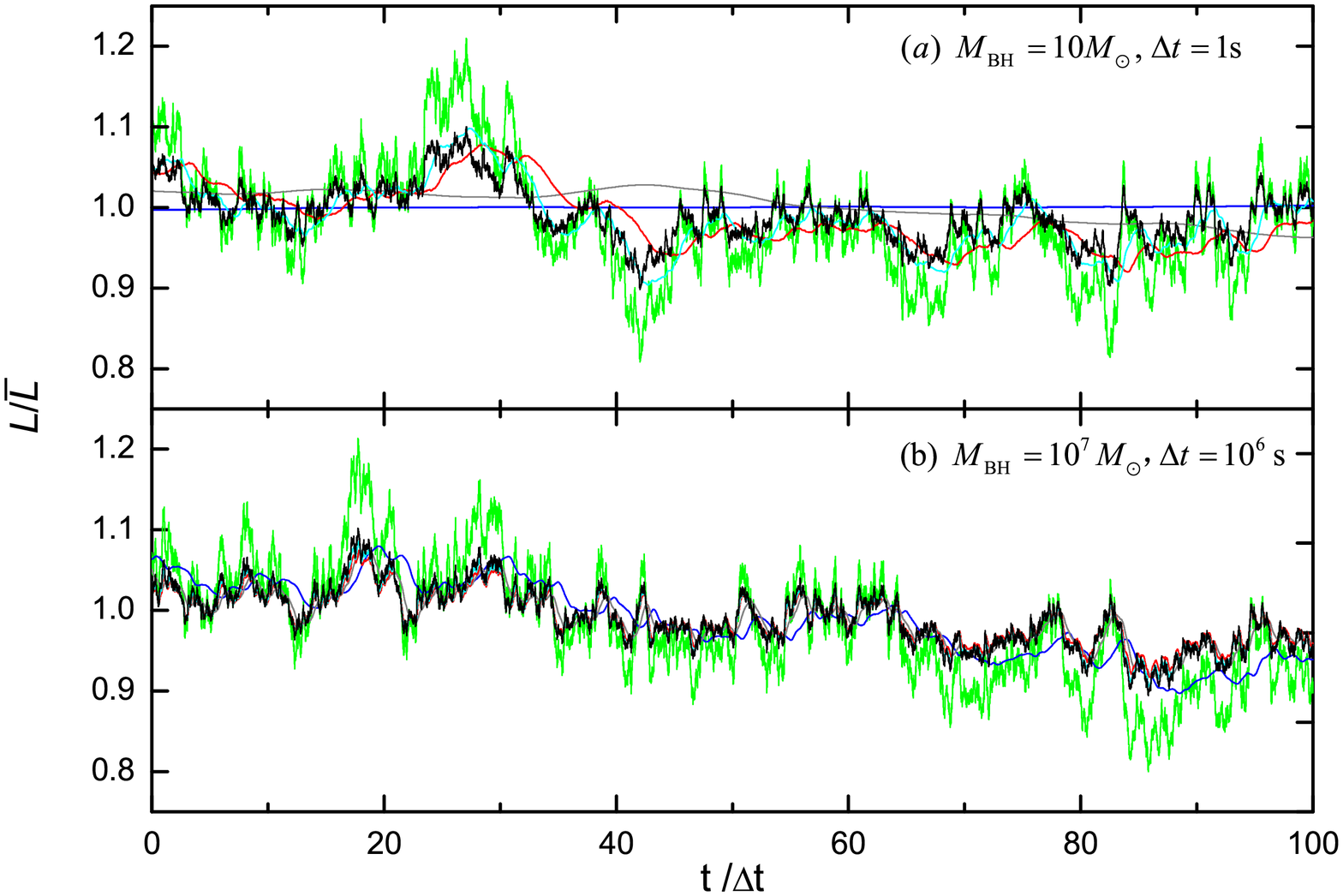}
\caption{Light curves of disk emission and jet emission with different black hole mass.
The top panel is for $M_{\rm BH}=10M_\odot$ and the bottom panel is for $M_{\rm BH}=10^7M_\odot$.
The blue, gray, red, and cyan curves represent the light curves of jet emission at frequency
$10^{10}{\rm Hz}$, $10^{11}{\rm Hz}$, $10^{12}{\rm Hz}$, and $10^{13}{\rm Hz}$, respectively.
The light curves of disk emission and $P_{\rm jet}$ are presented with green and black curves, respectively.}
\label{Fig:Light_Curve}
\end{figure}

Based on the models presented in Section~\ref{sec:Model},
we produce the light curves of disk emission and jet emission as shown
in Figure \ref{Fig:Light_Curve}.
The produced $L_{\rm X}/\,\overline{L_{\rm X}}$ and $P_{\rm jet}/\,\overline{P_{\rm jet}}$ are shown with green and black curves in this figure,
and the emission of jet $L_{\rm jet}(\nu)/\, \overline{L_{\rm jet}(\nu)}$
at different frequency $\nu$ is described with different color curves:
the blue curves are for $\nu=10^{10}{\rm Hz}$,
the gray curves for $\nu=10^{11}{\rm Hz}$,
the red curves for $\nu=10^{12}{\rm Hz}$,
and the cyan curves for $\nu=10^{13}{\rm Hz}$.
Here, $L_{\rm jet}(\nu)$ represents the luminosity of jet emission at frequency $\nu$ and
$\overline{L_{\{...\}}}$ ($\overline{P_{\rm jet}}$) represents the mean value of $L_{\{...\}}$ ($P_{\rm jet}$).
The light curves in the top (bottom) panel of this figure
are produced in the accretion system with $M_{\rm BH}=10M_\odot$ ($M_{\rm BH}=10^7M_\odot$)
in a duration of 100 ($10^8$) seconds,
where the total duration of simulation $T_{\rm tot}$ is around 1700 ($1.7\times 10^9$) seconds.
It can be found that the light curves of disk emission are
similar to those presented in the Figure~1 of \cite{Arevalo2006}.
Moreover, the amplitude of variability in the jet emission at low frequency, such as $\nu=10^{10}{\rm Hz}$,
may be weak compared with that in the jet power (disk emission).
However, the variability of jet emission at high frequency, such as $\nu=10^{13}{\rm Hz}$,
almost follows the variability of jet power.
Because of those behaviors, some interesting relations between disk emission and jet emission may be formed.
We should emphasize that the above behaviors are presented
in the accretion system with a constant $\overline{\dot{M}}$ (or $\overline{L_{\rm X}}$)
rather than an evolving $\overline{\dot{M}}$ (or $\overline{L_{\rm X}}$).
In the following part, we discuss these relations.

In order to reduce the data points in the top panel of Figure \ref{Fig:Light_Curve},
we divide the light curves with a bin size ($\Delta t$) of one second
and obtain the average luminosity in each bin as the new data.
We analyse the relation of disk emission and jet emission based on these new data.
The results are shown in Figure~\ref{Fig:Correlation},
which reveals the strong correlation between disk emission and jet emission.
In this figure, the relation between disk emission and
jet emission at different observed frequency $\nu$ is represented with different symbol.
The symbol ``$\vartriangle$'' is for the jet emission at frequency $\nu=10^{10}{\rm Hz}$,
``$\times$'' for $\nu=10^{11}{\rm Hz}$, ``$\circ$'' for $\nu=10^{12}{\rm Hz}$,
and ``$\star$'' for $\nu=10^{13}{\rm Hz}$.
In each panel of this figure, the solid line represents the logarithmic linear fitting result of the data,
and the dashed line represents the relation of the fundamental plane with a given black hole mass, i.e.,
$\log(L_{\rm jet}/\overline{L_{\rm jet}})\propto 0.60\log(L_{\rm X}/\overline{L_{\rm X}})$.
According to this figure, it is easy to find that the relation of disk emission and
jet emission at frequency $10^{10}{\rm Hz}$
almost runs parallel with the abscissa axis, i.e., the axis of disk emission.
This behavior is similar to that presented in Figure~5 of \cite{King2013},
which suggests that the results in this work may provide an explanation for this kind of peculiar correlations.
In addition, the slope of relation increases with increasing observed frequency of jet emission.
The largest slope of these relations is reached by the jet emission at frequency $10^{13}{\rm Hz}$ with a slope of $0.58$,
close to the slope of the fundamental plane relation.
This reveals that the relation of fundamental plane is only followed by the jet emission at frequency $10^{13}{\rm Hz}$
for the situation studied here, i.e., $M_{\rm BH}=10M_\odot$, $\Delta t=1{\rm s}$, and with a duration of 100~s observation.

Based on Figure~\ref{Fig:Light_Curve},
one may find that the amplitude of variability in the jet emission at the same frequency
(such as $10^{10}\rm Hz$) may be different for different black hole mass.
Owing to this fact, we show the relations of disk emission and jet emission ($\nu=10^{10}\rm Hz$)
for different black hole mass in Figure~\ref{Fig:Correlation_10}.
In this figure, the symbols ``$\star$'', ``$\circ$'',  ``$\vartriangle$'', and ``$\times$''
represent the data (i.e., averaged luminosities in time span $\Delta t$) from the accretion system with
$M_{\rm BH}=10^3M_{\odot}\,(\Delta t = 10^2\rm{s})$,
$3.6\times 10^5M_{\odot}\,(\Delta t = 3.6\times 10^4\rm{s})$,
$10^7M_{\odot}\,(\Delta t = 10^6\rm{s})$, and
$10^9M_{\odot}\,(\Delta t = 10^8\rm{s})$, respectively.
In each panel of this figure, the solid line represents the logarithmic linear fitting result of the data,
the dashed line represents the relation of fundamental plane with a given black hole mass,
and the data are from $10M_{\rm BH}/M_{\odot}$ seconds simulation ($T_{\rm tot}\sim 170M_{\rm BH}/M_{\odot}\;{\rm s}$).
According to this figure, the slope of relation between disk emission and jet emission increases
with increasing black hole mass, where $\Delta t \propto M_{\rm BH}$ is adopted.
The upper limit of the relation slope is $0.46$,
close to the slope of the fundamental plane relation.
In a practical context, the observations are usually performed in a short period (such as 1ks)
and luminosities used to estimate the relation of fundamental plane
are selected with an average spacing of several days.
In order to model this situation, we average luminosities in $\Delta t=1{\rm ks}$ as the new data
and use the data with a spacing time of 3 days to estimate the relation of disk emission and jet emission.
Figure~\ref{Fig:Correlation_10-1d} shows the relations for these luminosities,
where data are from $1.5\times 10^7{\rm s}$ simulation ($T_{\rm tot}\sim 1.2\times 10^8{\rm s}$).
In this figure, the symbols ``$\star$'', ``$\circ$'',  ``$\vartriangle$'', and ``$\times$''
represent the data from the accretion system with
$M_{\rm BH}=10M_{\odot}$,
$10^3M_{\odot}$,
$3.6 {\times} 10^5M_{\odot}$, and
$10^7M_{\odot}$, respectively.
Since the viscous timescale of the accretion disk around the black hole with $M_{\rm BH}\sim 10^{9}M_\odot$
is significantly larger than 1 hour, the accretion system with $M_{\rm BH}\sim 10^{9}M_\odot$ is not considered.
Figure~\ref{Fig:Correlation_10-1d} shows that the slope of relations decreases with increasing black hole mass.
The largest slope of these relations ($0.63$),
found in the accretion system with $M_{\rm BH}=10M_\odot$,
is close to the slope of fundamental plane.
That is to say the deviation of our relations from that of fundamental plane
becomes significant for systems with more massive black hole in the case studied in Figure~\ref{Fig:Correlation_10-1d}.
For the comparison with the observations, a dotted line,
i.e., $\log(L_{\rm jet}/\overline{L_{\rm jet}})\propto 0.06\log(L_{\rm X}/\overline{L_{\rm X}})$,
which describes the relation of disk emission and jet emission at $\nu=5\rm GHz$ found in NGC~4395 (\citealp{King2013}),
is added in the bottom left panel of this figure.
In this panel, the red ``*'' represents the data of $L_{\rm X}$ and $L_{\rm jet}(\nu=5\rm GHz)$
based on the models in Section~2.
The logarithmic linear fitting about these data presents a relation
of $\log(L_{\rm jet}/\overline{L_{\rm jet}})\propto 0.19\log(L_{\rm X}/\overline{L_{\rm X}})$,
which is described with a red solid line in this panel.
It is shown that our relation in the accretion system with $M_{\rm BH}=3.6 {\times} 10^5M_{\odot}$
is shallower than that of the fundamental plane, but steeper than that from observation.
This may imply that other mechanisms may contribute to produce the relation found in \cite{King2013}.

\subsection{A Simple Analysis}\label{sec:Analysis}
\begin{figure}
\plotone{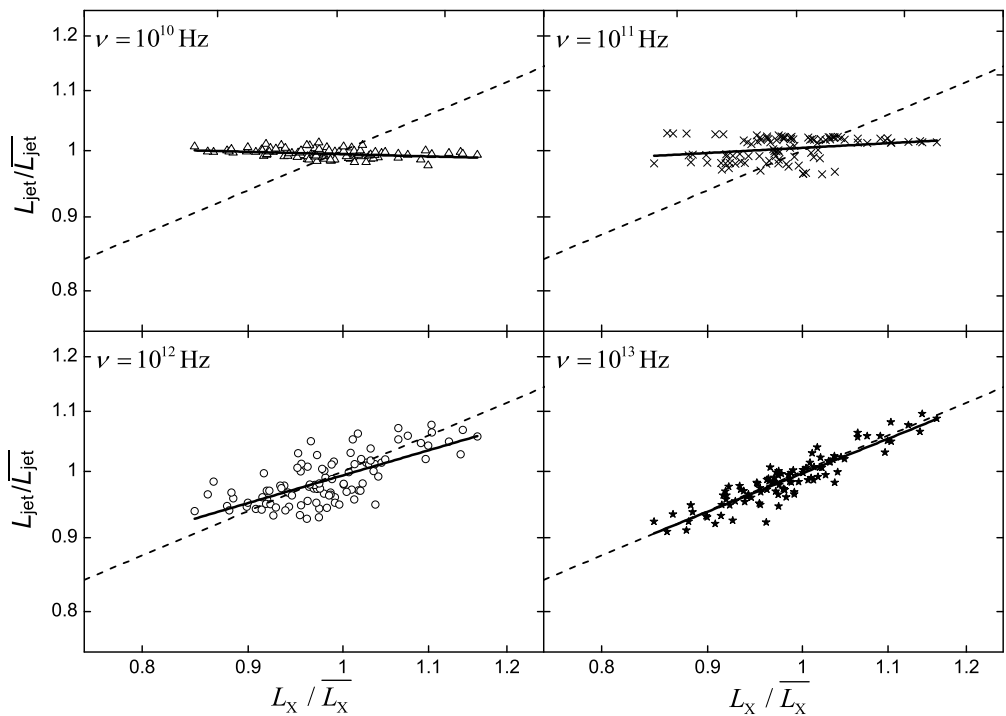}
\caption{Relations between disk emission and jet emission,
where $\vartriangle$, $\times$, $\circ$, and $\star$ represent
the relation with jet emission at frequency
$10^{10}{\rm Hz}$, $10^{11}{\rm Hz}$, $10^{12}{\rm Hz}$, and $10^{13}{\rm Hz}$, respectively.
The solid line and dashed line in each panel represent the logarithmic linear fitting result
and the relation of the fundamental plane with a given black hole mass, respectively.}
\label{Fig:Correlation}
\end{figure}

The results presented in Figure \ref{Fig:Correlation}
can be well understood on the basis of the light curves presented
in Figure \ref{Fig:Light_Curve}, especially those in the top panel.
Taking the jet emission at frequency $10^{10}\rm Hz$ as an example,
it is shown that the luminosity of jet emission at this frequency
keeps almost constant whereas the luminosity of disk emission exhibits strong variability.
Therefore, the relation corresponding to this frequency runs parallel with the axis of disk emission.
This behavior can be found in Figure~\ref{Fig:Correlation} for the data with the symbol ``$\vartriangle$'' .
The reason
for the correlation running as the symbol ``$\star$'' is as follows.
According to the models presented in Section~\ref{sec:Model},
the variability of disk emission and jet power mainly
depend on the fluctuation of accretion rate $\dot{M}$ and ${\epsilon}(r)$,
i.e., Equations (\ref{EQ:L_X}) and (\ref{EQ:Jet Energy}).
These two equations reveal that the variabilities of disk emission and
jet power are almost the same except for the amplitude of variability.
Furthermore, the variability of jet emission is modulated by the fluctuation of jet power.
Then, the variability of jet emission should be more or less in a way that works
as the variability of jet power and thus disk emission.
That is to say, a correlation between disk emission and jet emission should be naturally expected.
If the variability of jet emission well tracks the variability of jet power,
the variability of jet emission would follow that of disk emission except for the amplitude of variability.
This behavior can be found in the top panel of Figure~\ref{Fig:Light_Curve}
with cyan color light curve ($\nu=10^{13}\rm Hz$) or in the bottom right panel of Figure~\ref{Fig:Correlation}.

As discussed above, a correlation of disk emission and jet emission is naturally expected
owing to the fact that the jet power modulates the jet emission.
However, why the variability of jet emission at different frequency is different?
Since the jet emission at different frequency shares the same jet power,
the different behavior of variability should be only associated with the emissivity profile of these emission.
Figure~\ref{Fig:Jet_Emission_Profile} shows the emissivity profile of jet emission
for different observed frequency in the accretion system with $M_{\rm BH}=10M_\odot$,
i.e., $10^{10}{\rm Hz}$ (solid curve), $10^{11}{\rm Hz}$ (dashed curve),
$10^{12}{\rm Hz}$ (dotted curve), and $10^{13}{\rm Hz}$ (dot-dashed curve), respectively.
In this figure, ${I_\nu }(z,t)$ is the jet emissivity at frequency $\nu$ from location $z$ observed at time $t$.
Based on this figure, the luminosity of jet emission can be described as
\begin{eqnarray}\label{EQ:L_jet}
L_{\rm jet}(\nu,t) & \propto & \int_{z_0}^{z_{\rm out}}{{I_\nu }(z,t-(z_{\rm out}-z)\cos\psi/c)z\phi dz} \nonumber \\
 &\sim & \int_{{Z_{\rm{p}}} - \delta Z_p/2}^{{Z_{\rm{p}}} + \delta Z _p/2} {{I_\nu }(z,t-(z_{\rm out}-z)\cos\psi/c)z\phi dz},
\end{eqnarray}
where $z_{\rm out}$ is the outer location of jet emission,
$Z_{\rm p}(\nu)$ is the location of peak emissivity,
$\delta Z_p \sim W_{\rm log}Z_{\rm p}$,
and $W_{\rm log}$ is the logarithmic width of emissivity profile.
It can be expected that the variability of $L_{\rm jet}(\nu,t)$ in the timescale
$\delta t \lesssim \delta Z_p/{\upsilon _{{\rm{jet}}}}$ would be suppressed according to Equation (\ref{EQ:L_jet}).
This behavior has been reported in the observation of GX~$339-4$ (see Figure 3 of \citealp{Casella2010}
and Equation (15) of \citealp{Lin2012ApJ}).
As shown in Figure~\ref{Fig:Jet_Emission_Profile},
the larger value of $Z_{\rm p}$ corresponds to the lower value of observed frequency.
In addition, the value of $W_{\rm log}$ is almost the same for different observed frequency.
Then, the lower value of $\nu$ would cause a stronger suppression of the variability.
This can be found in Figure~\ref{Fig:Light_Curve},
where the variability of jet emission at low frequency (such as $10^{10}\;{\rm Hz}$)
is obviously weaker than those at high frequency (such as $10^{13}\;{\rm Hz}$).
This is the reason for the different variability behavior of jet emission observed at different $\nu$.

\begin{figure}
\plotone{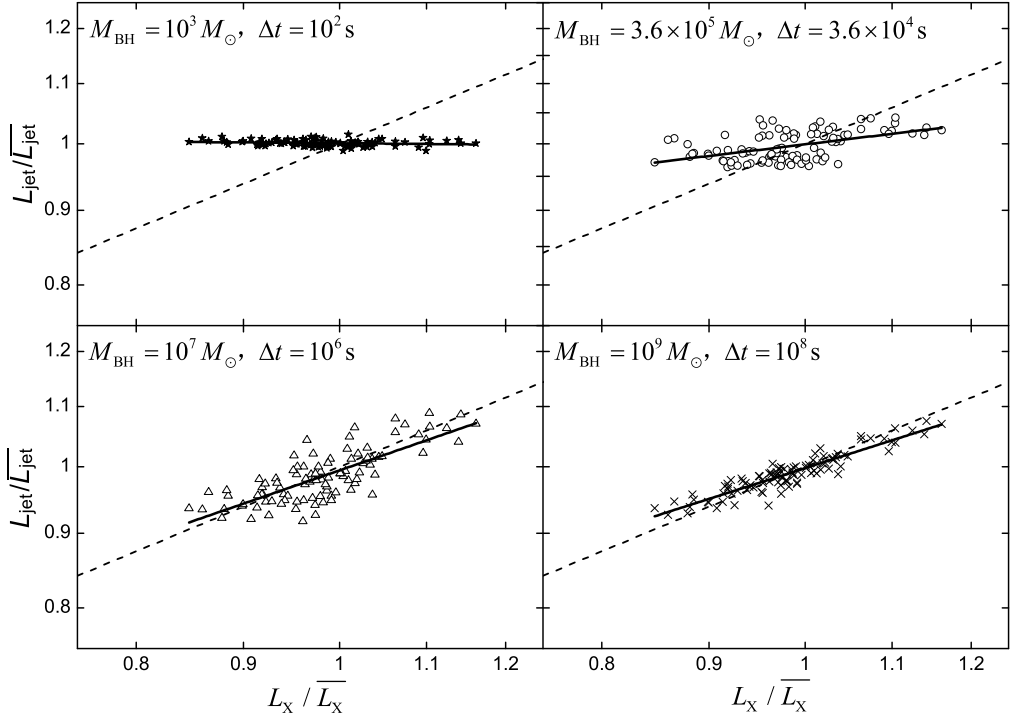}
\caption{Relations between disk emission and jet emission at frequency $10^{10}{\rm Hz}$,
where ``$\star$'', ``$\circ$'',  ``$\vartriangle$'', and ``$\times$''
are the data from an accretion system with
$M_{\rm BH}=10^3M_{\odot}\,(\Delta t = 10^2\rm{s})$,
$3.6\times 10^5M_{\odot}\,(\Delta t = 3.6\times 10^4\rm{s})$,
$10^7M_{\odot}\,(\Delta t = 10^6\rm{s})$, and
$10^9M_{\odot}\,(\Delta t = 10^8\rm{s})$, respectively.
The solid line and dashed line in each panel represent the logarithmic linear fitting result
and the relation of the fundamental plane with a given black hole mass, respectively.}
\label{Fig:Correlation_10}
\end{figure}

\begin{figure}
\plotone{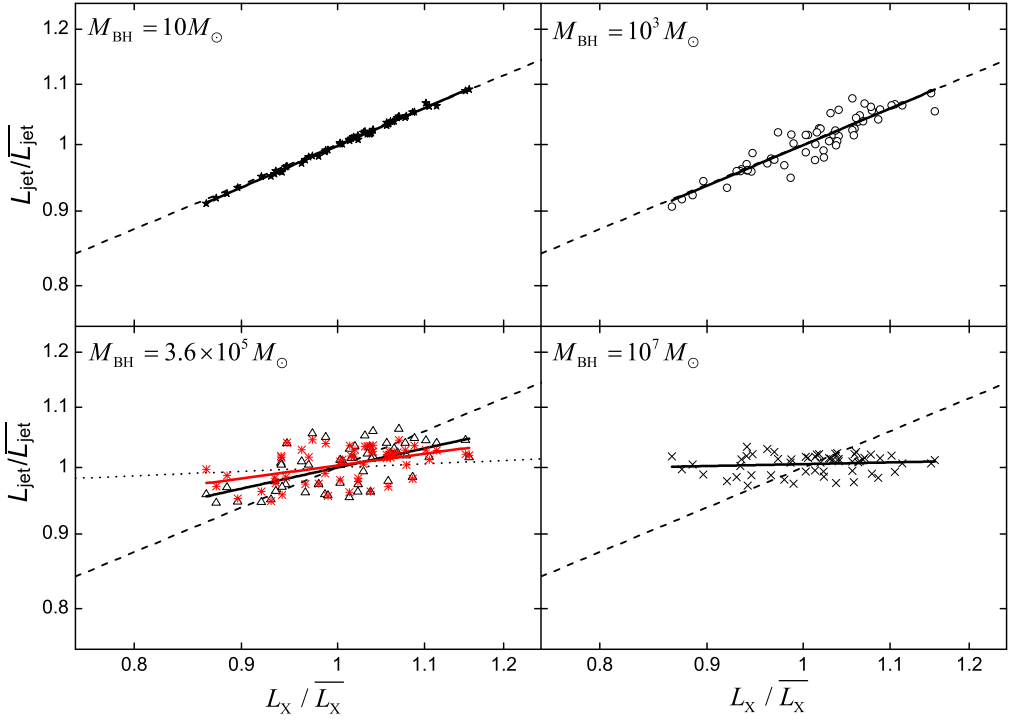}
\caption{Relations between disk emission and jet emission at frequency $10^{10}{\rm Hz}$,
where ``$\star$'', ``$\circ$'',  ``$\vartriangle$'', and ``$\times$''
are the data from an accretion system with
$M_{\rm BH}=10M_{\odot}$,
$10^3M_{\odot}$,
$3.6\times 10^5M_{\odot}$, and
$10^7M_{\odot}$, respectively.
The data are the averaged lumonisities in $\Delta t=1\rm ks$ with a spacing time of 3 days.
The solid line and dashed line in each panel represent the logarithmic linear fitting result
and the relation of the fundamental plane with a given black hole mass, respectively.
The red solid line represents the logarithmic linear fitting result about
the data with $L_{\rm jet}$ observed at $\nu=5\rm{GHz}$, which is shown with red ``*''.
The dotted line in the bottom left panel represents the relation of
$\log(L_{\rm jet}/\overline{L_{\rm jet}})\propto 0.06\log(L_{\rm X}/\overline{L_{\rm X}})$,
which was found in NGC~4395 (\citealp{King2013}).
\label{Fig:Correlation_10-1d}}
\end{figure}

\begin{figure}
\plotone{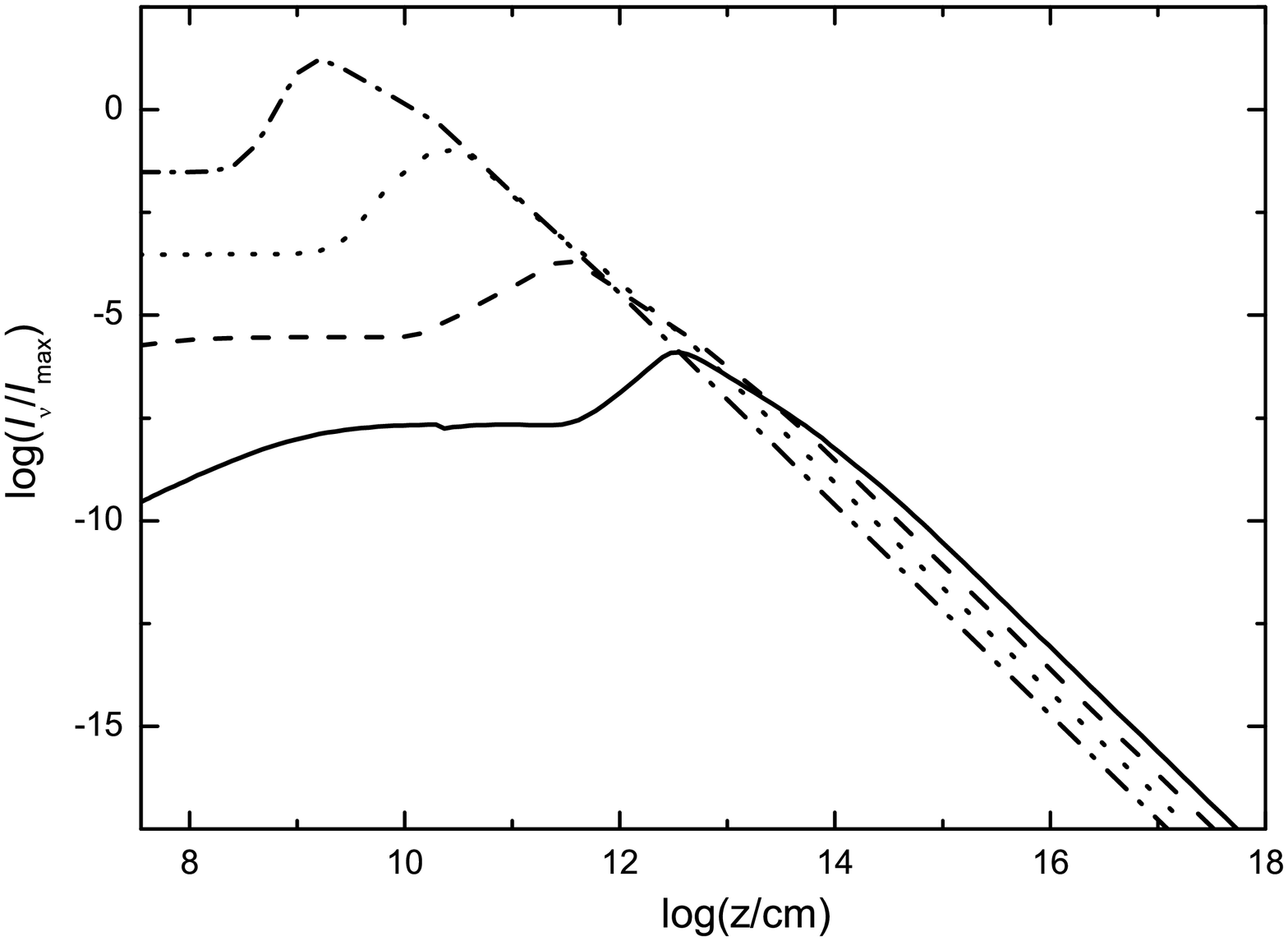}
\caption{Emissivity profile of jet emission at frequency
$10^{10}\,{\rm Hz}$ (solid curve), $10^{11}\,{\rm Hz}$ (dashed curve),
$10^{12}\,{\rm Hz}$ (dotted curve), and $10^{13}\,{\rm Hz}$ (dot-dashed curve), respectively.}
\label{Fig:Jet_Emission_Profile}
\end{figure}

Equation~(\ref{EQ:L_jet}) reveals that the variability of jet emission
in the long timescale $\delta t > \delta Z_p/{\upsilon _{{\rm{jet}}}}$
would follow that of $I_{\nu}$ and thus $P_{\rm jet}$ except for the amplitude of variability.
In the work of \cite{Heinz2003}, the relation of $L_{\rm jet}(\nu) \propto \dot{M}^{17/12}$ was derived
for typical flat-spectrum core-dominated radio jets.
Our simulations with a steady jet power confirm this relation but with a slightly different index for different $\nu$.
Then, one can expect that the relation of disk emission and jet emission
in the long timescale $\delta t > \delta Z_p/{\upsilon _{{\rm{jet}}}}$ would be
$L_{\rm jet}(\nu) \propto L_{\rm X}^{17/24}$,
where $L_{\rm X}\propto \dot{M}^{2}$ is adopted.
This behavior can be found in the top left panel of Figure~\ref{Fig:Correlation_10-1d}.
For the jet emission at $\nu=10^{10}{\rm Hz}$ from the accretion system with $M_{\rm BH}=10M_\odot$,
$\delta Z_p/{\upsilon _{{\rm{jet}}}}\sim 100 {\rm s}$ can be estimated
based on Figure~\ref{Fig:Jet_Emission_Profile}.
Then, the variability of jet emission with timescale $\delta t <100{\rm s}$ would be suppressed,
and that with timescale $\delta t > 100{\rm s}$ will follow the variability of $P_{\rm jet}$.
This is the reason for the different behaviors in the top left panel of Figure~\ref{Fig:Correlation}
and that of Figure~\ref{Fig:Correlation_10-1d}.
The increase of slope in Figure~\ref{Fig:Correlation_10} can also be understood as follows.
In Figure~\ref{Fig:Correlation_10}, $\Delta t \propto M_{\rm BH}$ is adopted.
However, the increase rate of $Z_P(\nu=10^{10}{\rm Hz})/\upsilon _{\rm{jet}}$ with $M_{\rm BH}$ does not follow that of $\Delta t$.
Then, the variability of jet emission in the accretion system with larger $M_{\rm BH}$ ($\Delta t$)
would present a better tracing behavior for the jet power compared to the case with lower $M_{\rm BH}$ ($\Delta t$).

As discussed in Sections~\ref{sec:Disk Variability} and \ref{Sec:Variability of Jet Power},
the value of $\gamma$ (or $\mu+s$) may affect the variability of disk emission (jet power)
in the timescale ${\delta t}\lesssim t_{\rm vis, in}$ (see Figure~2 of \citealp{Arevalo2006}),
where $t_{\rm vis, in}$ is the viscous timescale of ADAF in the inner region.
For the case in the present work, however, the timescale ${\delta t}\sim \delta Z_p/{\upsilon _{{\rm{jet}}}}$
(see Figure~\ref{Fig:Jet_Emission_Profile})
is significantly larger than $t_{\rm vis, in}$.
Then, the value of $\gamma$ (or $\mu+s$) may not have significant effect on the relations
presented in Figures~\ref{Fig:Correlation}-\ref{Fig:Correlation_10-1d}.

\section{Conclusions and Discussion}\label{sec:Conclusion}
The variability of accretion rate,
which is believed to induce the significant aperiodic variability of X-ray emission from disk,
may affect the energy injection of jet from accreting process.
Owing to this behavior,
correlations between X-ray emission from disk and jet emission may be formed
even if the mean value of X-ray luminosity keeps constant.
These correlations are found in this work.
It is shown that the correlation may be different from that of the fundamental plane.
The fitted results show that the slope of correlation may increase with increasing observed frequency of jet emission.
Taking an accretion system with $M_{\rm BH}=10M_\odot$ as an example,
it is shown that the slope of correlation varies from $0$ to $\sim 0.58$
for the jet observed frequency varying from $10^{10}\,\rm Hz$ to $10^{13}\,\rm Hz$,
where $L_{\rm X}\propto\dot{M}^2$ and total duration of observation is $100{\rm s}$.
For the luminosities spacing with 3~days,
which are used to estimate the fundamental plane in a practical context,
the slope of relation decreases with increasing black hole mass.
The shallow behaviour of our found relations is due to the suppression of variability
in the jet emission compared with that in the disk emission.
These results suggest that luminosities used to estimate the fundamental plane should cover an appropriate timescale,
in which the variability of jet emission is not significantly suppressed compared with that of disk emission.

The shallow relations found in this work may appear in the observations.
We compare the relations found in this paper with that of observations (\citealp{King2013}).
In the work of \cite{King2013}, a shallow relation,
i.e., $\log(L_{\rm jet}/\overline{L_{\rm jet}})\propto 0.06\log(L_{\rm X}/\overline{L_{\rm X}})$,
has been reported in the observation of NGC~4395.
In our work, we also find a relation,
i.e., $\log(L_{\rm jet}/\overline{L_{\rm jet}})\propto 0.19\log(L_{\rm X}/\overline{L_{\rm X}})$,
which is shallower than that of the fundamental plane but still steeper than that of \cite{King2013}.
The deviation of our result with that of observation may imply that
other mechanisms may contribute to produce the observed shallower relation.
Besides the low value of relation slope found in observations (\citealp{Bell2011}; \citealp{King2013}),
several sources, such as NGC 4051 (\citealp{King2011}), 3C 120 (\citealp{Chatterjee2009}), and 3C 111 (\citealp{Chatterjee2011}),
show negative relation between $L_{\rm X}$ and $L_{\rm R}$.
In these sources, the accretion rate is close to or larger than $0.01\dot{M}_{\rm Edd}$,
around which the transition from low/hard state to high/soft state appears in BHBs (e.g., \citealp{Yu2009}; \citealp{Dunn2010}).
Then, the negative relations may be due to the suppression of jet production in the high/soft state
(\citealp{Maccarone2003}; \citealp{Sikora2007}; \citealp{Coriat2011}; \citealp{Russell2011}).
This mechanism may work in NGC 4395, of which the accretion rate is $\sim 0.01\dot{M}_{\rm Edd}$.
The radiative efficiency may increase in the state transition from low/hard state to high/soft state (e.g., \citealp{Xie2012}).
In this case, relation of $L_{\rm X}$ and $L_{\rm jet}$ may be steeper than that of the fundamental plane.
This behaviour has been found in some BHBs and AGNs
(e.g., \citealp{Coriat2011}; \citealp{Corbel2013}; \citealp{Dong2014}; \citealp{Panessa2015}).
However, this mechanism could not explain the shallower relation found in \cite{King2013}
since it steepens the relation.
It should be noted that the models presented in Section~2 may be better applied to a source
in the low/hard state or high/soft state rather than in a transition state,
where the suppression of jet and the increase of radiative efficiency may appear.

\acknowledgments
We thank the anonymous referee for beneficial suggestions to improve the paper.
This work is supported by the National Basic Research Program of China (973 Program, grant No. 2014CB845800),
the National Natural Science Foundation of China (Grant No. 11403005, 11025313, 11222328, 11333004),
the Guangxi Science Foundation (Grant No. 2014GXNSFBA118004, 2013GXNSFFA019001),
the Project Sponsored by the Scientific Research Foundation of Guangxi University (Grant No. XJZ140331),
and the Key Scientific Research Project in Universities of Henan Province (Grant No. 15A160001).
%%%%%%%%%%%%%%%%%%%%%%%%%%%%%%%%%%%%%%%%%%%%%%%%%%%%%%%%%%%%%%%%%%%%%%%%%%%%%%%%%%%%%%%%%%%%%%%%%%%%%%%%%%%

%%%%%%%%%%%%%%%%%%%%%%%%%%%%%%%%%%%%%%%%%%%%%%%%%%%%%%%%%%%%%%%%%%%%%%%%%%%
%%%%%%%%%%%%%%%%%%%%%%%%%%%%%%%%%%%%%%%%%%%%%%%%%%%%%%%%%%%%%%%%%%%%%%%%%%%
\clearpage

\clearpage

\clearpage

\clearpage

\clearpage
\end{document}